\documentclass[sigconf]{acmart}
\usepackage{subfig}
\usepackage{float}
\usepackage{listings}
\usepackage{enumitem}
\usepackage{fancyhdr}
\usepackage{tikz}
\usepackage{setspace}
\newcommand*\circled[1]{\tikz[baseline=(char.base)]{
	            \node[shape=circle,draw,inner sep=0.6pt,color=white,fill=black] (char) {#1};}}

\usepackage[compact]{titlesec}
\titlespacing{\section}{0.2pt}{0.2ex}{0.2ex}
\titlespacing{\subsection}{0pt}{0.2ex}{0ex}
\titlespacing{\subsubsection}{0.1pt}{0.1ex}{0.1ex}
\setcopyright{none}
\renewcommand\footnotetextcopyrightpermission[1]{}
\AtBeginDocument{%
  \providecommand\BibTeX{{%
    \normalfont B\kern-0.5em{\scshape i\kern-0.25em b}\kern-0.8em\TeX}}}
\newcommand{\algcolor}[2]{\hspace*{-\fboxsep}\colorbox{#1}{\parbox{2.6in}{#2}}}
\definecolor{deepgray}{gray}{0.5}
\definecolor{lightgray}{gray}{0.8}

\newcommand{\algemphlg}[1]{\algcolor{lightgray}{#1}}

\usepackage[ruled,linesnumbered]{algorithm2e}
\begin{document}

\title{Introducing Fast and Secure Deterministic Stash Free Write Only Oblivious RAMs for Demand Paging in Keystone}


\author{Mriganka Shekhar Chakravarty}
\affiliation{%
 \institution{Indian Institute of Technology, Kanpur}
  \streetaddress{Indian Institute of Technology Kanpur, Kalyanpur, Kanpur - 208 016}
  \city{Kanpur}
  \country{India}}
\email{mriganka@cse.iitk.ac.in}

\author{Biswabandan Panda}
\affiliation{
 \institution{Indian Institute of Technology, Bombay}
  \streetaddress{Indian Institute of Technology Kanpur, Kalyanpur, Kanpur - 208 016}
  \city{Mumbai}
  \country{India}}
\email{biswa@cse.iitb.ac.in}

\begin{abstract}
Keystone is a trusted execution environment, based on RISC-V architecture. It divides the memory into a secure Keystone private memory and an unsecure non-Keystone memory, and allows code that lies inside the Keystone private memory to execute securely.  Simple demand paging in Keystone ends up leaking sensitive access patterns of Keystone application to the Operating System(OS), that is assumed to be malicious. This is because, to access the unsecure non-Keystone memory, Keystone needs support of the OS. To mitigate this, Keystone needs to implement oblivious demand paging while obfuscating its page access patterns by using Oblivious RAM(ORAM) techniques. This causes substantial slowdown in the application execution. 

In this paper, we bridge the performance gap between application execution time with unsecure and secure demand paging in Keystone by using Deterministic, stash free, Write only ORAM (DetWoORAM) for oblivious demand paging. We also show why DetWoORAM, that is a write-only ORAM, is sufficient for oblivious demand paging. DetWoORAM logically partitions the memory into a main area and a holding area. The actual pages are stored in main area. We propose two enhancements over DetWoORAM that improves the application execution slowdown. The first enhancement, which we call the Eager DetWoORAM, involves page preloading that exploits the deterministic access pattern of DetWoORAM, and tries to hide the ORAM latency. The second enhancement, which we call the Parallel DetWoORAM, involves spawning multiple threads and each thread performs a part of the DetWoORAM memory access algorithm. Compared to DetWoORAM that shows slowdown of [1.4x, 2x, and 3.24x], Eager DetWoORAM and Parallel DetWoORAM provide slowdown of [1.2x, 1.8x, and 3.2x] and [1.1x, 1.1x, and 1.4x], for k= 3, 7, and 15, respectively. 
\end{abstract}

\maketitle
\thispagestyle{fancy}


\fancyhead[C]{\fontsize{8}{10} \selectfont Appeared in CARRV'21 workshop (\href{https://carrv.github.io/2021}{\nolinkurl{https://carrv.github.io/2021}}), co-located with ISCA'2021}

\lhead{}
\rhead{}
\vspace{-5pt}
\section{Introduction} \label{Intro}
Keystone\cite{keystoneref}, a trusted execution environment based on RISC-V architecture, isolates the memory into secure Keystone private memory and unsecure non-Keystone memory or the unsecure DRAM, and allows Keystone application that lies inside the Keystone private memory to execute securely, even in presence of a malicious privileged software(e.g. Operating System(OS)). To run large applications, Keystone needs to undertake demand paging by utilizing the unsecure non-Keystone memory as the backing store, where the swapped out pages are hashed and stored in encrypted form. Even though the page tables are private to Keystone and page management is handled by the secure Keystone runtime, thereby eliminating the page fault side channel\cite{controlchannel}, the OS can still see the address of the pages that are loaded and evicted in and out of Keystone as it needs support of the OS to access the non-Keystone memory. Simple demand paging in Keystone thus leaks access pattern information to the untrusted OS at a page size granularity, and this has been shown to be unsecure by Xu et al\cite{controlchannel}.

\textbf{Page based side channel due to simple demand paging: }
Page access pattern and hence the control flow of applications can be leaked due to page loads, as well as page evictions in simple demand paging as shown below.\vspace{-2pt}
    
    \begin{lstlisting}[caption=Access Pattern Leak, label = pseudo] 
    int secret  =  input();   //Get input from user
    Page *P = (Page *)malloc(4*sizeof(Page));
    if (secret){
    A:    write(P[0], 0);     //Fill first page with all 0
          if (secret == 1){
    B:        write(P[1],0);  //Fill second page with all 0
    C:        write(P[2],0);  //Fill third page with all 0
          }
          else{
    D:        write(P[2],0);  //Fill third page with all 0
    E:        write(P[1],0);  //Fill second page with all 0
          }
    F:    write(P[0],0);      //Fill first page with all 0
    G:    write(P[3],0);      //Fill fourth page with all 0
    }
    else
    H:    write(P[1], 0);     //Fill second page with all 0
    \end{lstlisting}
    \indent \textit{Access pattern leak due to page load:}
    Let us consider the pseudocode \ref{pseudo}. Let us assume that the active memory of the process executing the above code does not contain Page P[0] and P[1]. The first \textit{if} statement decides if the line labeled \textit{A} or \textit{H} will be executed. Depending upon which line is executed, either page P[0] or P[1] will be loaded into the memory, which if the OS sees, can infer about the value of \textit{secret}.

    \indent \textit{Access pattern leak due to page eviction:}
    Page evictions are performed whenever there is no space in memory and we need to load a page to serve a page fault. Page evictions change the content of the memory, irrespective of the evicted page is dirty or not. This is because the evicted page is encrypted with an IND-CPA encryption scheme. Hence they are always visible to adversary. Let us consider the same pseudocode \ref{pseudo}, and assume that the active memory is of three pages and the page replacement policy used is LRU. Also let us assume that the first \textit{if} condition is \textit{True}. By the time line labeled \textit{F} finishes execution, the pages in the memory are P[0],P[1] and P[2], with P[0] being the most recently used page. Either of page P[1] and P[2] can be the least recently used page, depending upon the value of \textit{secret}. Now when the line labeled \textit{G} is executed, either page P[1] or P[2] has to be evicted to make place for page P[3] in the memory. If the OS sees which page is evicted, it can learn about the \textit{secret}.

    \indent \textbf{Prior Works:}
    The idea of using Oblivious RAM(ORAM) in demand paging was first proposed in Sanctum\cite{sanctum}. Aga and Narayanasamy introduced Oblivious Page Access Module(OPAM) in\cite{opam} for secure demand paging in Intel Security Guard eXtension(SGX). Their work is rendered unsecure by Roy et al in\cite{reusedist}. We implement OPAM as well as Path ORAM(PORAM), which inspired OPAM, in demand paging, and find that the average application performance slowdown with PORAM and OPAM are 27x and 7.4x respectively(refer Figure \ref{figure:ORAMtogether}).
    \\ \indent \textbf{Our goal} is to reduce the application performance slowdown of oblivious demand paging while maintaining the security guarantees that Keystone offers.
    
    \indent {\textbf{Our Observation:}}

    \fancyhead[L]{\fontsize{7}{9} \selectfont Introducing Fast and Secure Deterministic Stash Free Write Only Oblivious RAMs for Demand Paging in Keystone}
    \fancyhead[R]{\fontsize{7}{9} \selectfont CARRV Workshop’21, June 2021, Valencia, Spain}
    \chead{}

    We observe that write only ORAMs, instead of read-write ORAMs are sufficient to prevent access pattern leakage, provided we use the Keystone's UnTrusted Memory(UTM) as the backing store. This is because the Keystone can access UTM without the support of OS. Deterministic, stash free, Write only ORAM(DetWoORAM), a write only ORAM, logically partitions the memory into a main area and a holding area. The main area stores the actual physical pages, and the holding area stores the pages for as long as they are not unloaded into the main area. We also see that in DetWoORAM\cite{detwooram}, a write only ORAM, for every page write request, one page is written into the holding area and $K$(the ratio of size of main area to the size of holding area) pages are written into the main area. This is preceded by refreshing the cipher of these $K$ pages, meaning that these $K$ pages are read, decrypted, and re-encrypted. Addresses of the $K+1$ pages written by DetWoORAM, are deterministic and independent of the address of the actual page required to be written. 
    \\ \indent \textbf{Our Approach :} Based on this observation, we propose two enhancements over DetWoORAM. The first enhancement, which we call the Eager DetWoORAM(EDetWoORAM), involves preloading the $K+1$ pages into a \textit{preload buffer} and refreshing the cipher of these $K+1$ pages even before the actual page write request arrives. The second enhancement, which we call the Parallel DetWoORAM(PDetWoORAM) involves spawning multiple threads, and the threads concurrently refresh the cipher of the $K$ pages to be written into the main area, thereby decreasing the average slowdown of application execution. Oblivious demand paging with EDetWoORAM improves slowdown for usual values of K, i.e, 3, 7, 15 to [1.2x, 1.8x, and 3x] and PDetWoORAM improves it to [1.1x, 1.1x, and 1.4x] from [1.4x, 2x, and 3.24x] for oblivious demand paging with simple DetWoORAM, respectively. The slowdown is measured with respect to application execution time with simple demand paging that simply encrypts and hashes the page into the non-Keystone memory.
    

\section{Background and Motivation}    
\label{Background}
In this section, we provide a quick overview of Keystone components that are related to our work and also provide a background on DetWoORAM.

\subsection{Keystone}
The components of Keystone that are related to our work are:
\begin{figure}[t]
  \centering
  \includegraphics[width=\linewidth]{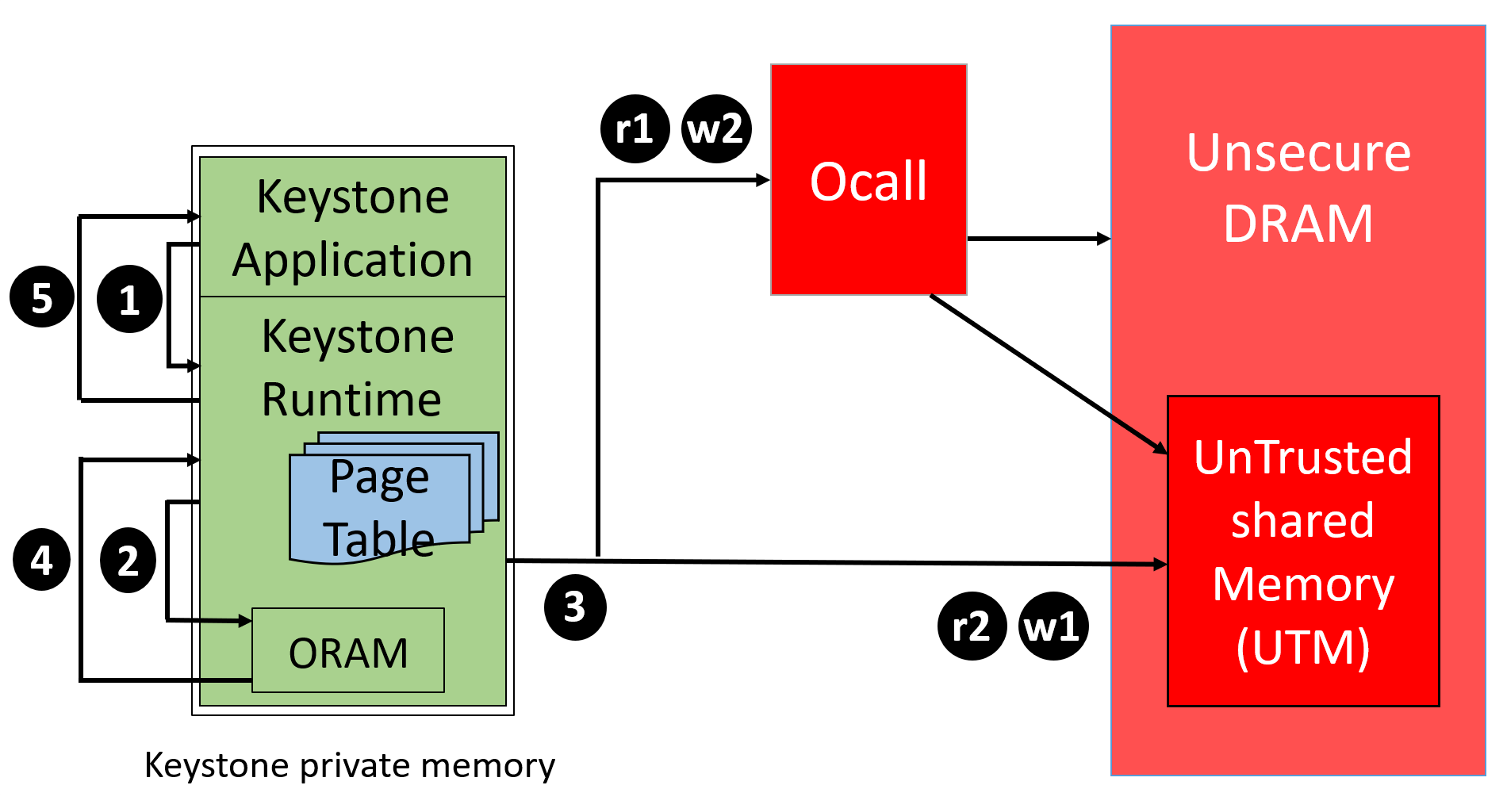}
  \vspace{-12pt}
  \caption{Demand paging using ORAM.}
  \Description{Steps involved in demand paging}
  \label{figure:steps_demand_paging}
\end{figure}

\begin{enumerate}[noitemsep,topsep=2pt, leftmargin = *]
    \item \textbf{OS:} The OS is assumed to be malicious and can tamper with memory locations in DRAM that are not \textit{protected} by Keystone.
    \item \textbf{Keystone Application:} Keystone application is the application that needs isolation from the untrusted (privileged/unprivileged) non-Keystone applications.
    \item \textbf{Keystone Runtime:} Keystone applications ideally do not take the services of the OS while executing, as the OS is assumed to be malicious. The runtime provides necessary substitutes for the services provided by the OS(e.g. page management).
    \item \textbf{Keystone private memory:} Keystone isolates the physical memory into a secure Keystone private memory, and the unsecure DRAM or the non-Keystone memory. The Keystone private memory consists of the Keystone application and the Keystone runtime.
    \item \textbf{UnTrusted shared Memory(UTM):} The runtime shares the UTM, that is in the unsecure DRAM, with the OS, and hence, the runtime can directly access the UTM. All exchange of data between the Keystone application and the unsecure non-Keystone memory happens using the UTM as an intermediary. As the malicious OS can also access this UTM, the UTM is unsecure.
    \item \textbf{Outbound call (ocall):} The runtime can not directly access the unsecure non-Keystone memory, other than the UTM. If it needs to access the same, it does so by making a function call, called \textit{ocall}, implemented in the untrusted host. The OS can snoop on the parameters that get passed on to \textit{ocall}, thereby rendering \textit{ocall} as unsecure, unless the passed parameters can be encrypted, which certainly can't be in case the parameters are addresses that the ocall needs to work on.
\end{enumerate}

Despite the page tables being private to Keystone and page management being done by the secure Keystone runtime, the OS is aware of the page addresses that are loaded in and evicted out of the Keystone. This is because, Keystone can access the unsecure non-Keystone memory only through \textit{ocall}. Thus, the access pattern of the Keystone is leaked at a page size granularity. This has been shown to be unsecure in \cite{controlchannel}. Hence we resort to oblivious demand paging and this is achieved by using an ORAM while loading and evicting pages in and out of the Keystone private memory when a page fault occurs, as shown in\cite{reusedist}.

\indent 
Figure \ref{figure:steps_demand_paging} shows the steps involved in oblivious demand paging, with ORAM. The steps are the same as mentioned in \cite{reusedist}. In step \circled{1}, a page fault occurs at an instruction in the Keystone application and the control is delegated to the runtime via the RISC-V exception delegation register. In step \circled{2}, the runtime checks if there are enough free pages that can be allocated or there is a need for page eviction. If a page eviction is required, then it invalidates the page table entry of the victim page(selected by the replacement policy). Next, the runtime invokes the ORAM memory access function for the chosen victim page and the page number that will be loaded. In step \circled{3}, if a page eviction is needed, then we write the victim page into the unsecure non-Keystone memory via the ORAM. This is followed by loading the required page from the unsecure DRAM via the ORAM. To achieve this obliviously, the ORAM makes a sequence of reads (r1 and r2) and writes (w1 and w2) from the unsecure DRAM. To read a page, the ORAM, which is implemented in the runtime, uses \emph{ocall} to copy the page \textbf{(r1)} from the non-Keystone memory into the UTM, and then in, the ORAM reads the page \textbf{(r2)} from the UTM into the Keystone private memory. For writing a page, the ORAM first writes it (\textbf{w1}) into the UTM (as UTM is the intermediary) and then uses ocall that copies this page (\textbf{w2}) from the UTM into the unsecure non-Keystone memory. Finally in step \circled{4}, the ORAM decrypts the loaded page in the Keystone private memory, and updates the page table entry corresponding to the page. The runtime returns the control to the Keystone application to execute the faulting instruction again in step \circled{5}.

\begin{figure}[t]
  \centering
  \includegraphics[width=\linewidth]{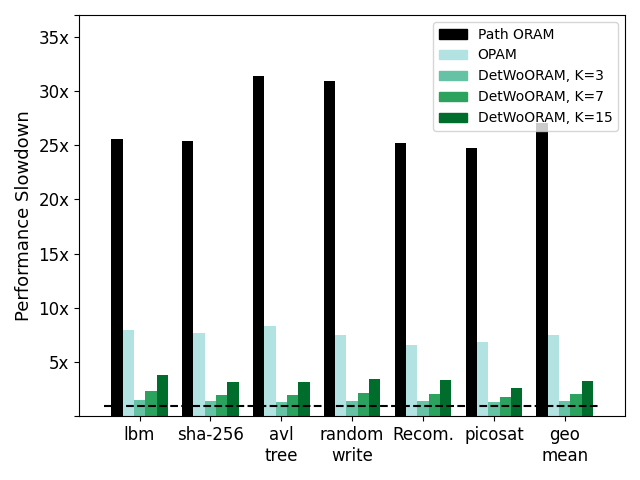}
  \vspace{-24pt}
  \caption{Performance slowdown with PORAM, OPAM and DetWoORAM, compared to simple demand paging with no ORAM(For benchmarks, refer section\ref{Evaluation}). Lower the better.}
  \vspace{-6pt}
  \label{figure:ORAMtogether}
  \Description{Performance slowdown with PORAM, OPAM and DetWoORAM. Lower the better.}
  
\end{figure}

\subsection{Deterministic, stash free, Write only ORAM}

Write only ORAMs relax the definition of security and leverage the relaxed definition to offer a faster ORAM. They consider reads to be oblivious, meaning that the adversary is not aware of the addresses from which data is read, but can see what addresses are being written at.

DetWoORAM is a write only ORAM introduced by Roche et al. in\cite{detwooram}. It views the unsecure memory as an array of pages and logically partitions the unsecure memory into a holding and a main area. The original paper proposes array of blocks of size $B$ bits, for some chosen B, we take the block size to be one page. DetWoORAM maintains a position map that keeps a record of where the latest copy of a particular page resides (main area or holding area). It also maintains pointers to holding area and main area, $Cur_H$ and $Cur_M$. The pointers tell where the next write access should happen in the respective areas. Assuming that the ratio of size of the main area to the size of holding area is K:1, the steps involved in a DetWoORAM are as follows:

\begin{enumerate}[noitemsep,topsep=2pt, leftmargin = *]
  \item \textbf{Write to holding area:} DetWoORAM encrypts and writes the required page into holding area at $Cur_H$, and updates the position map, reflecting that the latest copy of the desired page number is in the holding area at $Cur_H$, that gets incremented by one after the write. If $Cur_H$ exceeds the size of holding area, then DetWoORAM sets it to the start of holding area.
  
  \item \textbf{Refresh pages in the main area:} Each index in the array of pages that reside in the main area contains its corresponding page, that may or may not be the latest page. DetWoORAM performs a position map lookup and concludes where the latest copy of the next $K$ pages in main area are, starting from $Cur_M$. It reads the latest copies of these $K$ pages(reads are oblivious), decrypts, re-encrypts and writes them back into the main area at their corresponding addresses. Finally, DetWoORAM updates the position map of these $K$ pages and increases the $Cur_M$ pointer by $K$ and if $Cur_M$ exceeds the size of main area, then it sets $Cur_M$ to start of main area.
\end{enumerate}

\textbf{Motivation:}
The problem with oblivious demand paging is the application performance slowdown that comes with it. To provide the access pattern obliviousness, ORAMs need to perform a lot of spurious memory accesses and perform cryptographic operations such as encryption and decryption, which causes the performance slowdown. We implement oblivious demand paging with PORAM\cite{pathoram}, OPAM\cite{opam} and a DetWoORAM. We measure the slowdown of applications mentioned in section \ref{Evaluation} with respect to simple demand paging, i.e, simply encrypting and hashing the evicted pages without any ORAM and the same is shown in Figure \ref{figure:ORAMtogether}. We observe that the average slowdowns obtained from PORAM and OPAM are 27x and 7.4x, respectively, which is very slow, whereas DetWoORAM, being a write only ORAM, shows relatively better slowdowns with an average of 1.4x, 2x and 3.2x with K=3, K=7, and K=15 respectively.

\section{Write Only ORAM in Keystone} \label{woram_relevance}

Figure \ref{figure:steps_demand_paging} shows how the Keystone writes and reads, to and from the unsecure DRAM, through the UTM. Here we make an observation that, instead of using the UTM as an intermediary between the Keystone and the unsecure non-Keystone memory, we can utilize the UTM as the backing store for demand paging. The advantage of doing this is two-folds:

\begin{enumerate}[noitemsep,topsep=2pt, leftmargin = *]
    \item We can prevent double copying of pages, i.e., from keystone private memory to UTM and then from UTM to the unsecure non-Keystone Memory, and vice-versa.
    \item Given that Keystone runtime can directly access the UTM without seeking support from the OS, this renders the OS oblivious to Keystone’s read accesses to the UTM.
\end{enumerate}

Though the OS is oblivious to the read accesses of Keystone in UTM, the write accesses(page evictions) are still visible to the OS, as it can periodically take snapshots of the UTM and observe the changed data at various addresses, thereby learning where the writes have taken place. We discussed that this is unsecure in section \ref{Intro}. Exploiting this property of the UTM, we use Write Only ORAM in the Keystone runtime, while doing demand paging, using the UTM as the backing store, and the Keystone application demands for the required UTM size before application execution. If the UTM is full and the the Keystone application needs more memory, then the Keystone exits gracefully giving an out of memory error. To the best of our knowledge, Keystone does not support a UTM larger than 64MB, but as mentioned in Keystone\cite{keystoneref}, the size of Physical Memory Protection(PMP) regions can be as long as the entire DRAM region, and thus we envision future work in the expansion\footnote{Increasing the UTM in Keystone is supposed to be a new feature(Personal communication with Dayeol Lee, author of \cite{keystoneref}, dated 6th Jan, 2021).} of UTM size in Keystone.

\begin{figure}[t]
  \includegraphics[width=\linewidth]{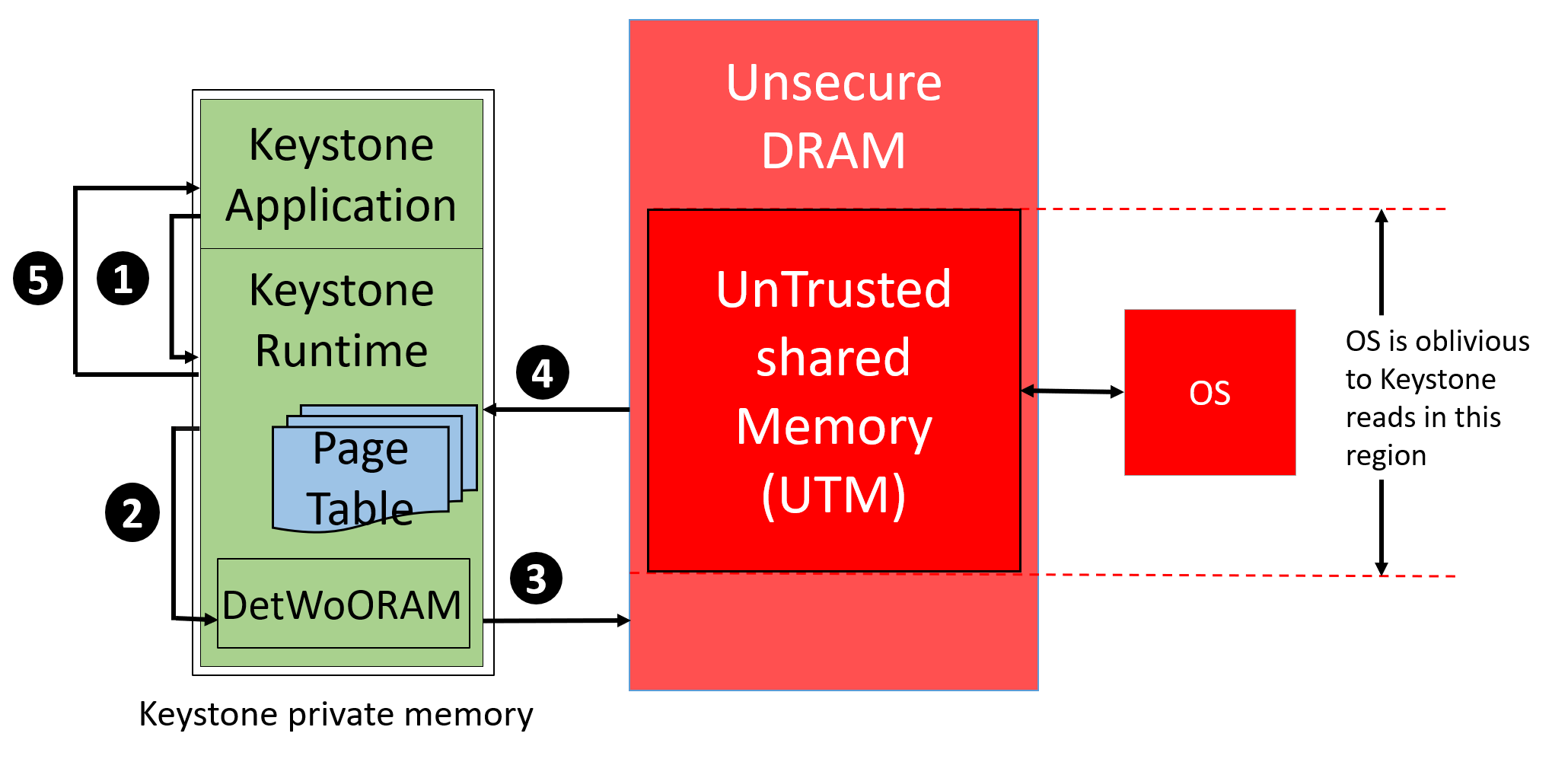}
  \vspace{-21pt}
  \caption{Write Only ORAM for Demand Paging.}
  \vspace{-3pt}
  \Description{Write Only ORAM for Demand Paging.}
  \label{figure:woram_block_diagram}
\end{figure}

We implement a DetWoORAM\cite{detwooram} that is a write only ORAM, in the Keystone's runtime, and Figure \ref{figure:woram_block_diagram} shows the steps involved in oblivious demand paging using DetWoORAM. It starts by delegating the control to the runtime fault handler via the RISC-V exception delegation registers, once a page fault occurs at an instruction in Keystone application(step \circled{1}). In step \circled{2}, the runtime checks if a page eviction needs to be performed and if needed, it chooses a victim page(chosen by the replacement policy) and invalidates its page table entry. In step \circled{3}, if a page needs to be evicted, the runtime writes the victim page into the UTM via the DetWoORAM. In step \circled{4}, the runtime loads the required page from the UTM with a normal read request, decrypts it, and validates the page table entry of the loaded page. Finally, in step \circled{5}, the runtime returns the control back to the Keystone application and the faulty instruction is executed again.

\section{Enhancements on DetWoORAM}
We propose two enhancements on DetWoORAM, Eager DetWoORAM (EDetWoORAM) and the Parallel DetWoORAM (PDetWoORAM). Both the enhancements discover a notion of parallelism that exist in DetWoORAM memory access algorithm. Our propositions of these enhancements are in no way specific to Keystone, and can be extended for use in any other trusted execution environment. In fact, we implement these enhancements independent of Keystone based on an x86 architecture with Linux. Keystone does not support multi-threading to the best of our knowledge, but it has the potential to support the same as mentioned in the Keystone\cite{keystoneref}. This motivated us to simulate our enhancements on Keystone. 

\subsection{EDetWoORAM} \label{EagerDet}
EDetWoORAM exploits the deterministic nature of DetWoORAM and preloads the relevant pages into the Keystone private memory before the actual write request arrives. 
\subsubsection{Observation : }
\indent In probabilistic ORAMs such as PORAM\cite{pathoram}, the ORAM memory access algorithm is probabilistic in nature, meaning that an ORAM tree path is accessed based on the mapping of the requested page to a leaf of the ORAM tree, and thus, the pages that are to be be accessed are discovered only after the actual page write(page eviction) request arrives. We make an important observation that in DetWoORAM, in contrast to PORAM, the ORAM access algorithm is completely \textit{deterministic} in nature, meaning that irrespective of the page that needs to be written, it is known precisely, which addresses(pages) in the memory need to be accessed.
\subsubsection{Design and Implementation : }
\indent We see that irrespective of which page is to be written, DetWoORAM makes $K$ writes in the next $K$ indices of the main area starting at $Cur_M$, and one write is made in the holding area at $Cur_H$, where $K$ is the ratio of the size of main area to the size of holding area. The idea with EDetWoORAM is to spawn a thread that \emph{preloads} the latest copy of these $K + 1$ pages from the UTM(i.e., from the holding or the main area) into the Keystone’s private memory even before an actual write request for page eviction arrives, and refresh them, i.e, decrypt, re-encrypt, and hash them. The thread works in the background. We put any write(eviction) request that arrives amid the preloading, into a \textit{blocking state} that is implemented by making the page fault handler loop, doing nothing as long as the background thread does the preloading, and exit the loop once the preloading ends.
\begin{figure}[t]
  \centering
  \includegraphics[width=\linewidth]{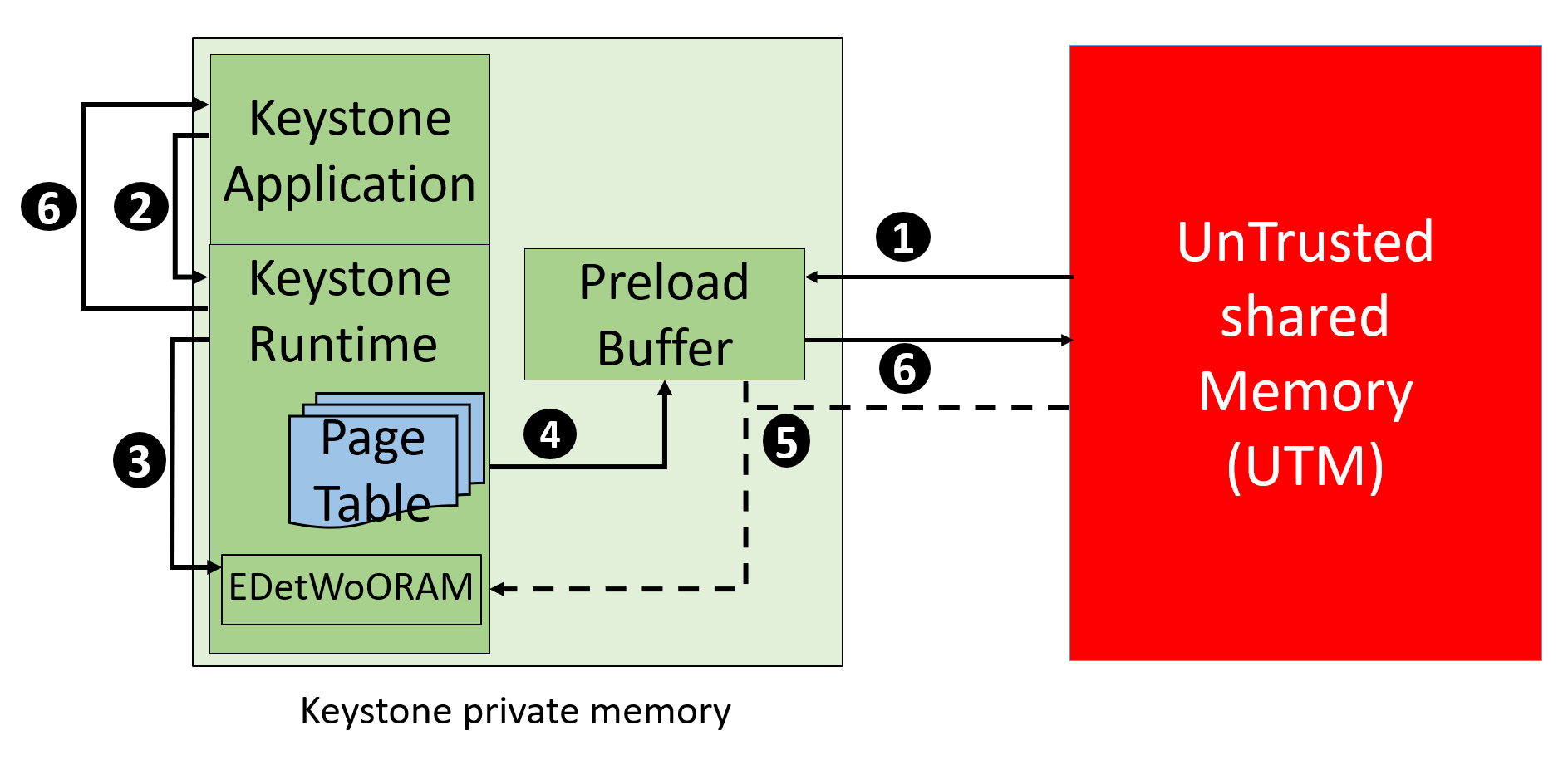}
  \vspace{-24pt}
  \caption{Steps in demand paging with Eager DetWoORAM.}
  \Description{Steps in demand paging with Eager DetWoORAM.}
  \label{figure:EagerSteps}
\end{figure}

For preloading, we reserve a buffer, called \textit{preload buffer} of $K + 1$ pages in the Keystone private memory, and this relaxes us of the responsibility to predict when a DetWoORAM access will be made and thus when to preload these $K+1$ pages. This is because these pages are preloaded, refreshed and stored in the buffer and become useful only when a DetWoORAM write request arrives. In classical client-server setting for ORAM, the \textit{preload buffer} can be introduced on the secure client side. 

Once an actual page write request arrives, the page is encrypted, hashed and written into the respective address in the \textit{preload buffer}, depending upon if it is in the range of the main area pages, or has to be kept in the holding area. We then spawn a thread again that unloads the \textit{preload buffer} back into the corresponding addresses of the UTM. An important thing to note is that, while preloading and unloading of the \textit{preload buffer}, we block all write requests as mentioned above to prevent data races. This does not cost us performance, as in the simple DetWoORAM access, we either way have to wait for the DetWoORAM write access algorithm to finish before proceeding with normal execution.

Figure \ref{figure:EagerSteps} shows the steps involved in demand paging with EDetWoORAM. In step \circled{1}, $K+1$ pages are preloaded into the \textit{preload buffer}, decrypted and re-encrypted. In step \circled{2}, once a fault occurs in the Keystone application, the control is delegated to the runtime fault handler via the RISC-V exception delegation register. In step \circled{3}, the runtime decides if a page eviction needs to be made, and if needed, it chooses the victim page(chosen by the replacement policy) and invalidates the page table entry of the chosen victim page. In step \circled{4}, if an eviction is needed, then the page to be evicted is written into the \textit{preload buffer}. In step \circled{5}, the faulting page is read into the Keystone Private memory from the UTM or the \textit{preload buffer} depending upon where the latest copy of the page is. Finally, in step \circled{6}, the \textit{preload buffer} is unloaded while the control is transferred back to the Keystone application.

\subsection{PDetWoORAM}\label{ParallelDet}

In this section, we introduce the Parallel DetWoORAM. Please refer to the algorithm \ref{par_woram_algo}. The algorithm is same as the DetWoORAM algorithm introduced in \cite{detwooram}.

\begin{algorithm}[t]
    \SetKwFunction{FMain}{WORAM\_WRITE}
    \SetKwProg{Fn}{Function}{:}{}
    //$p^{th}$ write at address a with data d\\
    \SetAlgoLined
    \Fn{\FMain{a, d}}{
        D[N + (p  mod  M)] = ENC(d) \\
        Pos[a] = N + (p mod M) \\
        s = $\lfloor p * K \rfloor mod N$ \\
        e = $\lfloor (p + 1) *  K \rfloor mod N$ \\
        //Refresh K main area blocks per-write\\
        \For{$i \in  [s,e)$}{
            \algemphlg{{D[i] = ENC( DEC( D[Pos[i]] ) )}}\\
            \algemphlg{Pos[i] = i}
        }
        p  = p + 1
    }
\caption{Parallel DetWoORAM algorithm (shaded loop is parallelized).}
\label{par_woram_algo}
\end{algorithm}

\subsubsection{Drawback of DetWoORAM : }
Let us say that the size of the main area in DetWoORAM is $N$ and the size of the holding area is $M$. The ratio of the size of main area to the size of holding area is K. In order to write a page into the back-store, a normal demand paging scheme needs to make one page write to the back store, which is the victim page. In contrast to this, demand paging with DetWoORAM makes $K+1$ writes, among which one of the pages is the victim page. The additional writes are the sources of obliviousness to the adversarial OS. In the best case, DetWoORAM makes one additional write meaning that the value of $K$ is one. We also know that the ratio of the size of the main area to the size of holding area directly determines what fraction of the entire ORAM memory is useful for storing actual data.

If $K$ is set to one, then half the memory, i.e, $1/(K+1)$, gets wasted in order to introduce obliviousness. Higher the value of $K$, better is the memory utilization, as for every one block in the holding area, we have more blocks in the main area, where the actual data can reside. But increasing the value of $K$ comes with a major drawback. This requires a higher number of actual writes, i.e, $K+1$ actual writes for one DetWoORAM write request. Hence, with increase in $K$, the slowdown for a DetWoORAM write request, and hence servicing a page fault increases linearly.

\subsubsection{Implementation : }
Considering Algorithm \ref{par_woram_algo}, which is same as the DetWoORAM algorithm in \cite{detwooram} the idea is to parallelize the shaded loop on line number 9 and 10. We see that the iteration of writes in the loop are independent of each other, meaning that the outcome of the iterations is invariant to the order in which the iterations are executed. We spawn $K$ threads and thread $i$ is responsible for performing the $i^{th}$ iteration of the loop. The improvement in slowdown due to the increase in number of threads, nullifies the effect of slowdown due to increase in $K$. We do not necessarily have to spawn $K$ threads. Given the liberty to spawn $T$ threads, thread $i$ performs the $p^{th}$ iteration of the shaded loop such that, \textit{p mod T = i} and \textit{p < K}.

\textbf{Summary:} As EDetWoORAM preloads the required pages into Keystone private memory, even before a page fault occurs, we can say that depending upon the rate of page fault, we can hide a significant fraction of the DetWoORAM latency. This is because, more the time in between two consecutive page faults, more time we get to preload the required pages into the \textit{preload buffer}. Hence EDetWoORAM for demand paging is a good fit in situations where the application runs with a working set large enough such that the rate at which faults occur is less. On the other hand, PDetWoORAM is suitable in situations where the processor is mostly idle, thereby giving us the liberty to spawn a large number of threads. PDetWoORAM gives aggressive performance improvements as compared to EDetWoORAM.


\section{EVALUATION}    \label{Evaluation}
\begin{table}[t]
 \begin{tabular}{||c | c||} 

 \hline
 \textbf{Parameter} & \textbf{Value} \\ [0.5ex]
 \hline\hline
 Page Size & 4kB \\ 
 \hline
 Page table entry size & 8B \\
 \hline
 UTM size & 64 MB \\
 \hline
 Levels of Page walk & 4 \\
 \hline
 Page encryption & chacha20 \\
 \hline
 Page Authentication & Hashed Message Authentication Code\\
 \hline
 Page replacement policy & FIFO \\ 
 \hline
 Working Set size & 15 pages(32kB)\\
 \hline
 Keystone Version & v0.3\\[1ex] 
 \hline
 Keystone Runtime & Eyrie\cite{keystoneruntime}\\[1ex] 
 \hline
 
\end{tabular}
\caption{Keystone parameters.}
\vspace{-15pt}
\label{tab:param}
\end{table}

We used Keystone on RISC-V QEMU\cite{keystonecode} to emulate all our experiments. Table \ref{tab:param} shows the parameters that are related to our evaluation. We used six applications to test our ORAMs. We run these applications with a very small active set of just 15 pages. This is so as to encourage a large number of page faults which in turn tests the performance of our page fault handler. Since we need to fit the applications into the UTM, the working set sizes have been kept small. The parameters of these applications are tuned so that their total working set size fits into the small UTM memory at our disposal. The applications that we use are:

\begin{itemize}[noitemsep,topsep=2pt, leftmargin = *]
    \item \textbf{random\_writes}: It’s a simple application that we wrote that makes 10k writes at random indices of a array of pages of length 1k. We use this application as representative of write intensive applications.
    \item \textbf{Sha\_256\cite{sharef}}: This application computes the sha-256 of bytes worth 1024 pages. We use this as representative of cryptographic primitives and we use this application from the Keystone's source code.
    \item \textbf{Avl\_tree\cite{avlcode}}: This application makes 100k insertions and deletions in an avl tree. We use it as a representative of commonly used data structures.
    \item \textbf{Picosat}: Picosat is a solver for propositional satisfiability\cite{picoref}. We use this application as representative of popular verification tools used today.
    \item \textbf{lbm}: This fluid dynamics application is taken from SPEC CPU 2017\cite{specref}. We use it as a representative of real world simulation applications from the scientific community.
    \item \textbf{Recommender\cite{recoref}}: This is a product recommendation engine built using collaborative filtering. This application is used as a representative of the popular machine learning frameworks.
\end{itemize}
\begin{figure}[t]
  \centering
  \includegraphics[width=\linewidth]{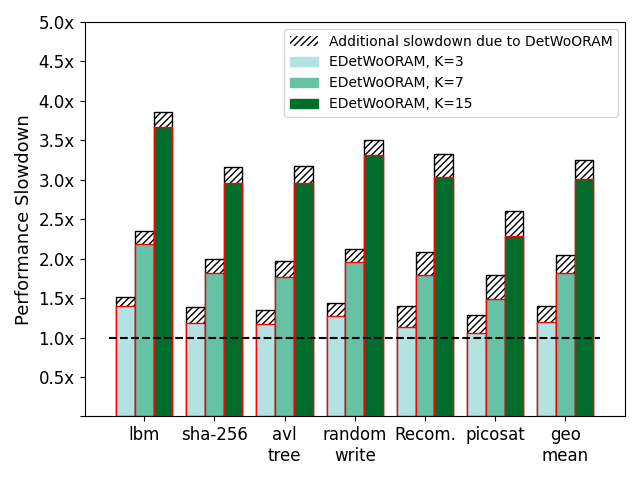}
  \vspace{-8mm}
  \caption{Performance slowdown with demand paging with Eager DetWoORAM compared to no ORAM in demand paging.}
  \vspace{-4pt}
  \Description{Performance slowdown with demand paging with Eager DetWoORAM compared to no ORAM in demand paging.}
  \label{eager_speedup}
\end{figure}

\textbf{Methodology:}
We use a makeshift evaluation strategy to measure the performance of demand paging with and without ORAMs. Some applications take too long to complete their execution. To get around this problem, we first run our applications with PORAM in demand paging in region of interest for 15 minutes, or till the end of execution, which ever is earlier. We count the number of page faults that are encountered, and thus get served in this time period. We then run the applications without any ORAM and with other ORAMs for exactly the same number of page faults. The time taken to serve the same number of page faults with and without various ORAMs give us a good idea regarding the performance of these ORAMs. The ratio of time taken to serve these many page faults with ORAM to without ORAM provides us the slowdown multiplier for that particular ORAM.

To simulate the PDetWoORAM, we run the applications normally with the DetWoORAM, and measure the amount of time that the application spends in DetWoORAM(say $t$) and also the total Keystone application execution time(say $T$). We implement the PDetWoORAM natively on a Linux based system in C language with \textit{pthread}\cite{pthread} library and observe the speedup that we get for the PDetWoORAM memory access algorithm(say $c$). We scale down the time spent in ORAM when run in Keystone, and thus arrive at our new/simulated execution time($T^{'}$). We calculate $T^{'}$ as $T^{'} = T - t + t/c$.


We simulate the EDetWoORAM in Keystone. In order to do so, we measure the time spent by Keystone in between every two consecutive page evictions(say $t$). We also measure the time taken by the ORAM to refresh the main area(say $T$). If $t < T$, then we discount $t$, otherwise, we discount $T$ from the total time spent in execution of the Keystone application. This is to simulate the fact that, while Keystone is busy not servicing page fault, we preload the required pages into the \textit{preload buffer}. \\[-7pt]

\textbf{Performance:}
Figure \ref{eager_speedup} shows the slowdown of application execution time with EDetWoORAM in demand paging and also with DetWoORAM in demand paging. We observe that the slowdown with EDetWoORAM is marginally reduced to 1.2x, 1.8x and 3x from 1.4x, 2x and 3.24x for K=3, K=7 and K=15 respectively as compared to demand paging with DetWoORAM.

\begin{figure}[t]
  \centering
  \includegraphics[width=\linewidth]{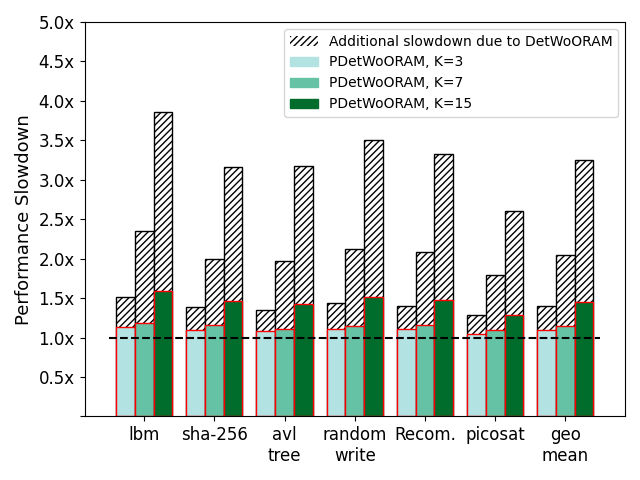}
  \vspace{-8mm}
  \caption{Performance slowdown with demand paging with Parallel DetWoORAM compared to demand paging with no ORAM.}
  \vspace{-6pt}
  \Description{Performance slowdown with demand paging with Parallel DetWoORAM compared to demand paging with no ORAM.}
  \label{woram_parworam}
\end{figure}

Finally in Figure \ref{woram_parworam} we show the performance slowdown for PDetWoORAM and DetWoORAM. We used $K$ threads for PDetWoORAM, where $K$ is the ratio of the size of main area to the size of the holding area. Increasing the number of threads beyond $K$ would not give us any more performance, as the DetWoORAM makes $K$ writes in the main area. We observe that, increasing the value of $K$ has small, but negligible effect on the performance of the PDetWoORAM. This is because the added workload due to increased $K$ gets distributed among all the threads. We also measured the overhead of creating $K$ threads, and saw that the thread creation time is 2.10\%, 4.49\% and 6.36\% of PDetWoORAM memory access time with K=3, K=7 and K=15, respectively. We find that on average slowdown of application execution time in demand paging with PDetWoORAM is substantially reduced to 1.1x, 1.1x and 1.4x from 1.4x, 2x and 3.24x for k=3, k=7 and k=15, respectively as compared to demand paging with simple DetWoORAM. We get such aggressive speedup solely by distributing the task of refreshing the K pages in main area to the K threads that we spawn. Each one of the thread refreshes, i.e., reads, decrypts, re-encrypts the page followed by appending to it, a hash of the page.

\section{Conclusion}
\label{Conclusion}
We saw how a write only ORAM can be used to do demand paging in Keystone. We introduced two enhancements to DetWoORAM, the Eager DetWoORAM and Parallel DetWoORAM. We see that EDetWoORAM provides marginal improvement of application execution slowdown and is a good candidate for oblivious demand paging if the rate of page fault is small. PDetWoORAM provides massive improvement of application execution slowdown(almost the best possible) and is a good fit for oblivious demand paging given the liberty to aggressively spawn threads.

\begin{acks}
This paper is in the loving memory of late Dr. Pramod Subramanyan. We would like to thank Supriya Suresh and Nirjhar Roy for helping with their experience in Keystone. We would also like to thank Yashika Verma and Anuj Mishra for reviewing our paper and point out incoherencies that might have crept into the paper at various stages.

\end{acks}

\bibliographystyle{ACM-Reference-Format}
\bibliography{acmart}
\appendix
\end{document}